# Effective Constitutive Relations for Simulating $CO_2$ Capillary Trapping in Heterogeneous Reservoirs with Fluvial Sedimentary Architecture

Naum I. Gershenzon[1], Robert W. Ritzi Jr.[1], David F. Dominic[1], Edward Mehnert[2]

[1]Department of Earth and Environmental Sciences, Wright State University, 3640 Col. Glenn Hwy., Dayton, OH 45435, USA

[2]Illinois State Geological Survey, Prairie Research Institute, University of Illinois at Urbana-Champaign, 615 East Peabody Drive, Champaign, IL 61820, USA

Corresponding author: Naum I. Gershenzon,

Tel: +1 937-775-2052,

e-mail: naum.gershenzon@wright.edu

**In the article several formulae should be corrected; see erratum to this article at the end.**

**Abstract**

Carbon dioxide ($CO_2$) storage reservoirs commonly exhibit sedimentary architecture that reflects fluvial deposition. The heterogeneity in petrophysical properties arising from this architecture influences the dynamics of injected $CO_2$. We previously used a geocellular modeling approach to represent this heterogeneity, including heterogeneity in constitutive saturation relationships. The dynamics of $CO_2$ plumes in fluvial reservoirs was investigated during and after injection. It was shown that small-scale (centimeter to meter) features play a critical role in capillary trapping processes and have a primary effect on physical- and dissolution-trapping of $CO_2$, and on the ultimate distribution of $CO_2$ in the reservoir. Heterogeneity in saturation functions at that small scale enhances capillary trapping (snap off), creates capillary pinning, and increases the surface area of the plume. The understanding of these small-scale trapping processes from previous work is here used to develop effective saturation relationships that represent, at a larger scale, the integral effect of these processes. While it is generally not computationally feasible to represent the small-scale heterogeneity directly in a typical reservoir simulation, the effective saturation relationships for capillary pressure and relative permeability presented here, along with an effective intrinsic permeability, allow better representation of the total physical trapping at the scale of larger model grid cells, as typically used in reservoir simulations, and thus the approach diminishes limits on cell size and decreases simulation time in reservoir simulations.

# 1. Introduction

One of the strategies being evaluated for reducing the atmospheric accumulation of greenhouse gasses is to capture carbon dioxide ($CO_2$) and inject it into geologic reservoirs for permanent sequestration (e.g., IPCC, 2005, Gale et al. 2015). Assessment of this strategy includes computational studies of how reservoir heterogeneity, over a range of scales, affects processes governing the dynamics, distribution, trapping, dissolution, mineralization, and ultimate fate of injected $CO_2$. Injection of $CO_2$ is also used for enhanced oil recovery (e.g., Ampomah et al. 2016; Dai et al. 2016; Soltanian et al. 2016). Here the focus is on the dynamics, distribution, and capillary trapping of $CO_2$ in the permeable part of the reservoir.

As reviewed by Krevor et al. (2015), a significant body of evidence, including results from laboratory studies, computational studies, and from field pilot injection tests, now indicates that residual trapping in the permeable part of the reservoir will be a primary mechanism for physically immobilizing $CO_2$ until it dissolves and mineralizes. Capillary trapping processes can be expected create residual $CO_2$ saturations of 20 to 30 percent in the permeable part of the reservoir; residual $CO_2$ that will not reach structural seals (e.g., shale cap-rock). Reservoirs without structural seals are now being considered in some inventories of U.S. storage capacity. Krevor et al. (2015) reviewed the pore-scale process of snap-off trapping within this context, and how it is represented in constitutive relationships through the hysteresis in capillary pressure and relative permeability as a function of phase saturation. One of their conclusions was that the influence of natural rock heterogeneity on residual trapping processes should be further investigated.

Here we consider natural rock heterogeneity associated with fluvial sedimentary architecture, as found in a number of candidate $CO_2$ reservoirs (Fig. 1). Recent work has shown how this type of sedimentary architecture can influence the residual trapping process in the permeable section of the reservoir (Gershenzon et al. 2014, 2015, 2016a, b, 2017; Trevisan et al. 2017a and b). Fig. 1 shows how fluvial bar deposits comprise sets of relatively finer- and coarser-grained cross strata (FG and CG rock types hereafter). In fluvial reservoirs such as the Lower Mt. Simon (Illinois, USA), these differences in grain size are the primary influence of variability in intrinsic permeability (Ritzi et al., 2016) within the permeable section of the reservoir. Fig. 2 is a model for these cross strata used in our previous work. As discussed in previous descriptions of this model, at 24% the CG cross sets percolate in 3-D (i.e., connect along tortuous pathways across any opposing boundaries of the domain) though this is not evident on the 2-D faces of the model. Connectivity is mostly vertical across a single unit bar, and tortuous laterally branching connections occur at the scale of assemblages of unit bars within a compound bar. Note also that the cross strata dip downward in the direction of paleoflow. The nature of these connections is important in the context of residual trapping. In addition to snap-off trapping, the sedimentary architecture creates capillary pinning. Though the FG cross sets are permeable relative to cap-rock seals and other larger-scale strata, their lower permeability relative to CG strata within the overall bar deposits enhances residual trapping within the larger bar deposits.

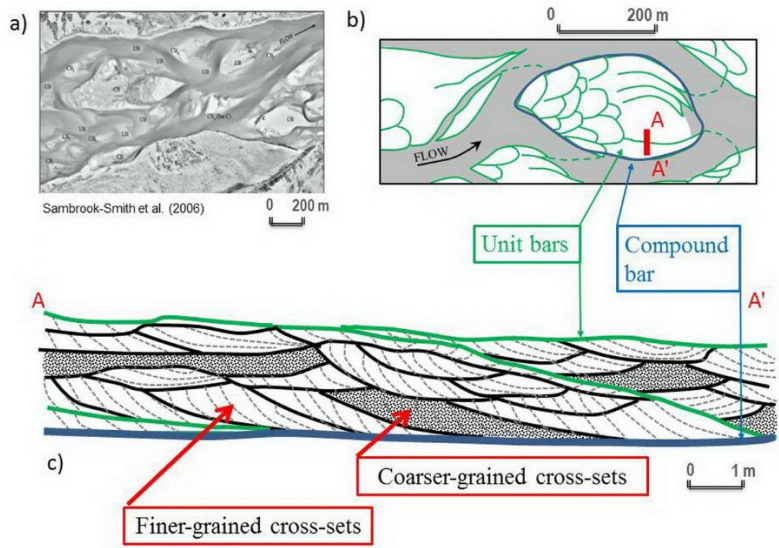

**Fig. 1**. Reservoir rock originally deposited by fluvial processes contains compound bar deposits (blue line), that comprise unit bar deposits (green line) that, in turn, comprise sets of finer- and coarser-grained cross strata. In some such reservoirs, such as the Lower Mt. Simon (Illinois, USA), these differences in grain size are the primary influence of variability in intrinsic permeability (Ritzi et al., 2016).

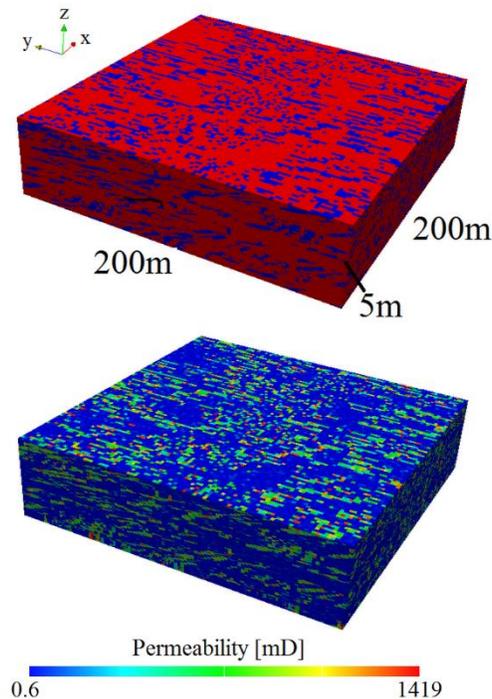

**Fig. 2** The image on the top shows the distribution of FG (red) and CG (blue) cross sets in the simulated reservoir, and the panel on the bottom shows the corresponding permeability. The cross sets are organized in unit bar deposits which are, in turn, organized within compound bar deposits. See Gershenzon et al. (2015) for images showing the larger-scale stratification. Here, in imaging

permeability, only the coarser grained, more permeable cross sets are readily discerned, because of the orders-of-magnitude larger permeability. Vertical exaggeration is 10×

As shown in Fig. 3, the heterogeneity inside of bar deposits creates a significant amount of residual trapping which includes this capillary pinning. Consider a CG cross set as shown in Fig. 4, with thickness $\xi$. Because of downward dip and complex, tortuous connectivity, along with vertical rise of buoyant $CO_2$, the local thickness is relevant to trapping. $CO_2$ has preferentially entered the CG cross set because of the relatively higher permeability and lower entry pressure through CG pathways. It will not buoyantly rise up into the overlying FG cross strata unless a critical capillary pressure is exceeded:

$$P_{CG}^{cr} = P_{e,FG} - \xi \Delta \rho g \qquad (1)$$

where $\Delta \rho g \xi$ is the buoyant force upward per area within this cross set; $\Delta \rho = \rho_w - \rho_{CO2}$; $\rho_w$ and $\rho_{CO2}$ are the density of brine and supercritical $CO_2$, respectively; $g$ is the gravitational constant; and $P_{e,FG}$ is the capillary entry pressure of $CO_2$ for the FG rock type. Fig. 5 illustrates Eq. [1]. The region included in the white square in Fig. 3 shows that at the saturation occurring within the CG cross sets, $P_{CG}^{cr}$ is not exceeded, and thus significant capillary pinning is occurring in the CG cross sets in reservoir simulations. The larger body of work studying this process suggests that this result will be common (Gershenzon et al. 2016b) over a wide range of scenarios.

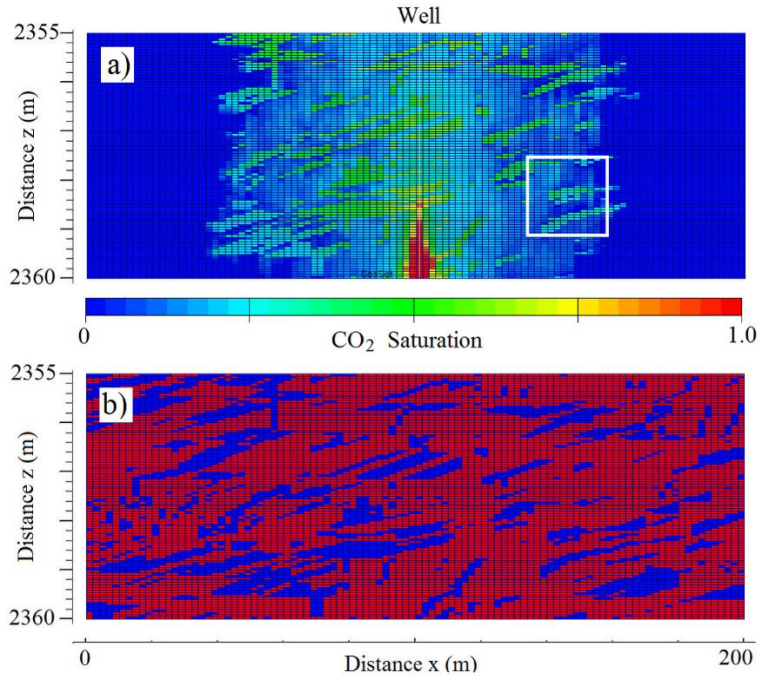

**Fig. 3** (a) Spatial distribution of CO$_2$ saturation in the middle cross-section of a reservoir after 250 days. The white square illustrates the (b) spatial distribution of CG rocks (blue) and FG facies (red) in the middle cross-section of the reservoir.

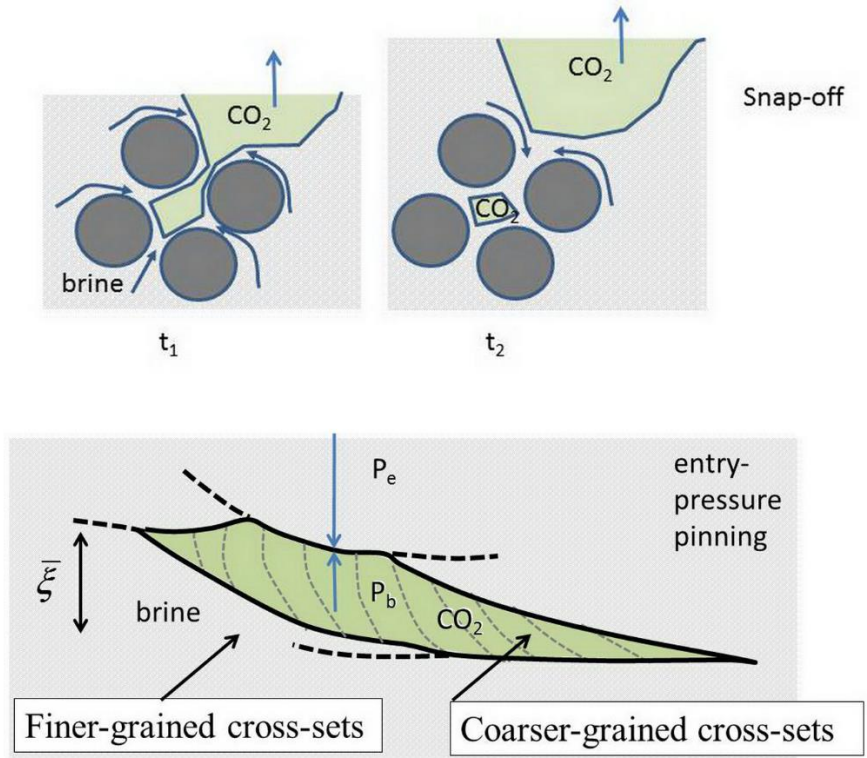

**Fig. 4**. Two residual capillary trapping processes. (a) snap-off trapping at the scale of pores. (b) entry pressure pinning at the scale of one of the coarser-grained cross-sets shown in Figure 1. $P_b$ is the buoyant pressure of the CO$_2$, and $P_e$ is the entry pressure of the FG rock type, which is less than $P_e$, causing capillary pinning within the CG rock type.

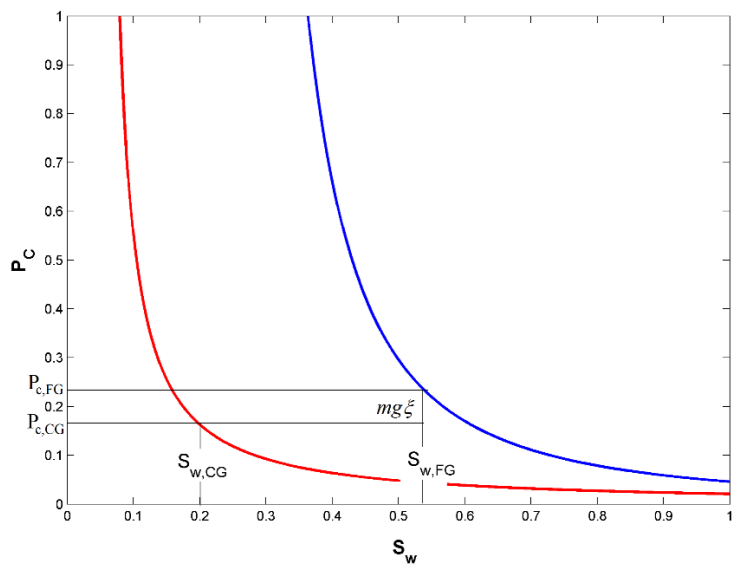

**Fig. 5.** The schematic figure of capillary pressure (*P*c) for FG (blue) and CG (red) facies as a function of water saturation ($S_w$) illustrating Eq. [1].

Though ignoring this small-scale heterogeneity in reservoir simulations leads to erroneous prediction, it is not computationally feasible to include these heterogeneities in simulations of realistically sized reservoirs. Indeed, the number of cells required is $10^9$ or more to simulate $CO_2$ injection in reservoirs with a size of $10^3$ to $10^5$ m in the lateral directions and from $10^1$ to $10^2$ m in the vertical direction. Our experience is that when full heterogeneity and hysteresis in saturation functions are represented (e.g., 12 non-linear constitutive relationships are required for just two textural rock types), it is not possible to achieve iterative convergence with existing simulators such as ECLIPSE, STOMP and TOUGH2 with cell numbers above a million or so. The goal of this work is to construct *effective constitutive relations for the homogeneous reservoir*, which will produce the same integrated characteristics as from more detailed simulations of a heterogeneous reservoir, such as the amount of capillary trapped $CO_2$ (including both snap-off and capillary pinning), the amount of dissolved $CO_2$, the amount of mobile $CO_2$, and give the general larger-scale shape and position of the plume. The idea is to create an effective intrinsic permeability tensor and an effective constitutive relationship for capillary pressure and relative permeability appropriate for the scale of larger bar deposits, avoid having to represent the smaller-scale features separately, make iterative convergence easier, and facilitate growing the size of the problem. We require that simulations using the effective parameters/relationships give approximately the same amount of trapping, from both snap-off trapping and capillary pinning, as is known to occur from the finer-scale simulations that represent finer- and coarser grain cross strata within the bar deposits. Our goal is to develop an approach that avoids generic fitting parameters, but rather is physically based, and uses quantifiable physical attributes for the CG and FG rock types comprised by the bar deposit, including their proportions, characteristic lengths, and individual petrophysical properties.

Previous work on effective parameters has not met all these objectives. The work of Saadatpoor et al. (2011) is the first attempt to implement upscaling accounting for trapping by snap-off and capillary pinning (residual trapping and local capillary trapping in their terms). However, they did it in the framework of a single-phase system which is suitable for the upscaling of intrinsic permeability but, obviously, not for the relative permeability and capillary pressure curves. Upscaling for a medium with capillary heterogeneity for the two-phase system applied to the $CO_2$ sequestration problem has been also initiated (e.g. Rabinovich et al. 2015; 2016 and references therein). Based on the approach described by Saadatpoor et al. (2011), Behzadi and Alvarado (2012) developed a two-phase flow upscaling by incorporating spatial connectivity to describe $CO_2$ upward migration in the capillary limit for a 2D heterogeneous reservoirs. Capillary pressure and buoyancy were taken into account. They showed their upscaling methodology yields improved the accuracy when percolation (spatial connectivity) was considered. Rabinovich et al. (2016) developed a method for calculating effective relative permeability curves in $CO_2$-brine systems during drainage, under steady state and capillary-limit assumptions. The analytical expressions for the effective parameters were derived. Although this method includes the effect of capillary pressure heterogeneity it doesn't model capillary trapping.

This previous work has not been based on the conceptual model for capillary pinning brought to light in our more recent work. Here we present an approach that is physically based on that understanding, and involves physically quantifiable statistics for the smaller scale-architecture, with no fitting parameters. We will suppose that a fluvial reservoir includes two types of cross strata, i.e. FG and CG rocks with different intrinsic permeability and, which is most important, with different capillary pressure and relative permeability curves. Thus, the task of obtaining effective petrophysical parameters includes 1) calculation of an effective anisotropic intrinsic permeability and 2) "averaging" the capillary pressure and relative permeability curves for $CO_2$ and brine for drainage and imbibition. In contrast to the approaches described above, we will consider a bimodal distribution of intrinsic permeability and, respectively, use two sets of saturation curves, i.e. for FG and CG rocks. The approach is presented in the next section. This approach is then used in an example implementation in Section 3, and results and conclusions are presented in Sections 4 and 5.

## 2. Approach

We begin by defining an averaging volume large enough to include both FG and CG rock types, as for example in the coarse grid region outlined by the square in Fig. 3a. Assume that within the volume the $CO_2$ density is uniform within CG and FG rock types. Mass is then conserved when defining a volume averaged, effective brine saturation $S_w^{eff}$:

$$S_w^{eff} = S_{w,CG} r_{CG} + S_{w,FG} r_{FG}, \tag{2}$$

where $r_{FG}$ and $r_{CG} = 1 - r_{FG}$ are the volume fractions of the FG and CG rock types, respectively. The brine saturation cannot become less than an effective irreducible water saturation of:

$$S_{wi}^{eff} = r_{CG} S_{wi,CG} + r_{FG} S_{wi,FG}. \tag{3}$$

The main idea behind this approach is to account for pinning within this averaging volume. Let $\bar{\xi}$ be a representative average thickness of CG strata (Ritzi et al. 2004) suggest the geometric mean thickness would be appropriate, based on typical thickness distributions quantified in nature). Consider that injected $CO_2$ has preferentially entered the CG rock type, and has potential to buoyantly rise in the vertical direction. The critical averaged capillary pressure in the CG rock type, that would have to be exceeded before $CO_2$ would enter the FG rock type above it, is:

$$P_{CG}^{cr,eff} = P_{e,FG} - \bar{\xi} \Delta \rho g$$

which occurs at an effective $CO_2$ saturation in the CG rock of $S_{CO2,CG}^{cr,eff}$. With $S_{CO2,FG} = 0$ and $S_{w,FG} = 1$, the effective critical $CO_2$ saturation for the entire averaging volume is then:

$$S_{CO2}^{cr,eff} = r_{CG} S_{CO2,CG}^{cr,eff}$$

and effective critical brine saturation is:

$$S_w^{cr,eff} = r_{CG} S_{w,CG}^{cr,eff} + r_{FG}$$

Given the above definitions of effective saturation, our first conjecture is that a reasonable approximation for the effective capillary pressure within the cell is given by the following averaging equations. The equation for drainage is

$$P_c^{eff}(S_{wi}^{eff} < S_w^{eff} < S_w^{cr,eff}) = r_{FG} P_{c,FG}(S_{w,FG}) + r_{CG} P_{c,CG}(S_{w,CG})$$
$$P_c^{eff}(S_w^{cr,eff} < S_w^{eff} < 1) = r_{FG} P_{e,FG} + r_{CG} P_{c,CG}(S_{w,CG})$$
(4)

and for imbibition is

$$P_c^{eff}(S_{wi}^{eff} < S_w^{eff} < S_w^{eff,im}) = r_{FG} P_{c,FG}(S_{w,FG}) + r_{CG} P_{c,CG}(S_{w,CG}),$$
(5)

where $S_w^{eff,im} = (1 - S_{CO2,FG}^{max}) r_{FG} + (1 - S_{CO2,CG}^{max}) r_{CG}$, $S_{CO2,FG}^{max}$ and $S_{CO2,CG}^{max}$ are the maximal residual saturations for the FG and CG facies.

What remains is to define an effective intrinsic permeability tensor, and effective relative permeability curves for the averaging volume. Considering the potential for vertical rise of the buoyant plume, the effective vertical permeability is of primary consideration. For the interval $S_w^{cr,eff} < S_w^{eff} \leq 1$, the effective $CO_2$ relative permeability $k_{CO2}^{eff}(S_w^{eff}) = 0$ and the effective brine relative permeability is $k^{eff}{}_w(S^{eff}{}_w) = k_{w,CG}(S_{w,CG})$. Our second conjecture is that below $S_w^{CR,eff}$ the effective relative permeabilities for $CO_2$ and brine can be well approximated by an averaging scheme consistent with how the effective intrinsic permeability is defined. There is abundant literature on computing an effective intrinsic permeability tensor. For perfectly layered strata with vertical to horizontal anisotropy in each layer, the effective permeability is given by the weighted arithmetic and harmonic means in lateral and vertical principal directions, respectively. For more complex architecture, without perfect layering, as in the case of higher-permeability facies embedded within lower permeability facies, a number of approaches have been proposed as reviewed for example by Rubin (2003) and Hunt and Idriss (2009). Hunt and Idriss (2009) evaluated the approaches in the context of correlated, bimodal distributions of hydraulic conductivity. None of the approaches were entirely satisfactory. The self-consistent approach (Dagan 1989; Rubin 2003) performed the best at proportions of coarser-grained rock around 30%, as in the case we investigate below. This self-consistent approach can be adapted for anisotropic media (Dagan 1989), and was shown to approximate effective conductivities in bimodal formations (Rubin 1995) and multimodal formations (Dai et al., 2004; 2005). With only relatively modest contrast in the intrinsic permeabilities between FG and CG rock types used here, or higher, and with the geologic structure considered here, the anisotropic permeability tensor defined by the self-consistent approach is essentially equal to that defined by the proportion weighted harmonic mean in the vertical direction and arithmetic mean laterally. Either approach could be adopted here. Because they give the same values, we adopt the simpler approach for the example here, and the intrinsic permeability tensor is given by:

$$k = \begin{pmatrix} k_x & 0 & 0 \\ 0 & k_y & 0 \\ 0 & 0 & k_z \end{pmatrix},$$ (6)

where $k_x = k_{CG}(\frac{h}{h+H} + \frac{H}{h+H}\frac{k_{FG}}{k_{CG}})$, and $k_z = k_{FG}\frac{h+H}{hk_{FG}/k_{CG}+H}$, where $h$ and $H$ are the average size of the CG and FG strata in $z$ direction. There is anisotropy not only between the vertical and lateral directions but also between the paleoflow ($x$) and perpendicular to paleoflow ($y$) directions. We assume that the ratio $k_x/k_y$ is proportional to the ratio $l_x/l_y$ between the typical length of strata in the $x$ and $y$ directions.

The effective relative permeability below $S_w^{cr,eff}$ is given in the vertical direction by:

$$k_{CO2}^{eff}(S_w^{eff}) = k_{CO2,FG}(S_{w,FG}) \cdot \frac{1}{r_{FG} + r_{CG}k_{CO2,FG}(S_{w,FG})/k_{CO2,CG}(S_{w,CG})},$$ (7)

Eq. [7] is applicable for imbibition as well.

The same approach is used to construct the relative permeability curve for water:

$$k_w^{eff}(S_{wi}^{eff} < S_w^{eff} < S_w^{eff,cr}) = k_{w,FG}(S_{w,FG}) \cdot \frac{1}{r_{FG} + r_{CG}k_{w,FG}(S_{w,FG})/k_{w,CG}(S_{w,FG})}$$

$$k_w^{eff}(S_w^{eff,cr} < S_w^{eff} < 1) = \frac{1}{r_{FG} + r_{CG}/k_{w,CG}(S_{w,CG})}.$$ (8)

Note that effective relative permeability should be a tensor. Here we will consider and use only the vertical component of this tensor, because most of simulators accept only scalar relative permeability.

## 3. Implementation, Example Application

*3.1 Fluvial-type reservoir*

Digital models that reproduce hierarchical architecture in fluvial channel-belt deposits have been developed (Ramanathan et al. 2010; Guin et al. 2010; Hassanpour et al. 2013). Based on digital models, geocellular (i.e., discretized) reservoir models have been created for computational experiments to study multiphase flow in fluvial-type reservoirs. Using the methodology described in Ramanathan et al. (2010), we generated reservoir realizations that contain two textural facies: FG cross-sets (76% of total volume) and CG cross-sets (Fig. 2). The permeability of each cell is assigned from the appropriate lithotype distribution (Ramanathan et al. 2010; Ritzi 2013). Table 1 includes the mean sizes of CG and FG clusters in all directions. These data will be used to calculate the effective averaged coefficient.

**Table 1.** The size of CG and FG clusters in *x*, *y* and *z* directions

| $h$ - mean size of CG clusters, $z$ direction in cell number | $H$ - mean size of FG clusters, $z$ direction in cell number | $l_x$ - mean size of CG clusters, $x$ direction in cell number | $l_y$ - mean size of CG clusters, y direction in cell number |
|---|---|---|---|
| 2.8 | 8.2 | 1.5 | 2.5 |

*3.2 Simulation setting*

The reservoir size was 200 m × 200 m × 5 m (1 million cells of size 2 m × 2 m × 0.05 m). Our focus is on evaluating the effective parameter relationships alone. There are many other issues outside the scope of this article, and already addressed by others, with respect to the grid upscaling problem, including the effects of numerical dispersion and representation of the injection well and boundary conditions when larger grid-blocs are used. Therefore, we choose to use the same fine-resolution grid spacing here, as used in the fine-scale simulations presented previously (Gershenzon et al. 2014, 2015, 2016a, b, 2017). That way, we can use the same injection scenarios and evaluate the results directly in terms of how well the effective parameters and relationships reproduce the total amount of residual trapping. The $CO_2$ was injected at a rate of 0.5 kg/s (0.01136 kg-mol/s) for 50 days into the bottom of a vertical well placed in the middle of reservoir at a depth of 2360 m. The pressure and temperature at the top of the reservoir are 230.8 bar and 53°C, respectively, such that the injected $CO_2$ was supercritical. The boundary conditions were that (1) there was no flow through the top and the bottom of the reservoir and (2) a Carter-Tracy aquifer (Carter and Tracy 1960) was applied for all others reservoir boundaries.

We used the commercial reservoir simulator ECLIPSE-300 with the CO2STORE option. Three components were included in simulations: $H_2O$, $CO_2$, and NaCl with initial total phase mole fractions of 0.9109, 0.0, and 0.0891, respectively. The water compressibility and viscosity were 4.35·10⁻⁵ 1/bar and 0.813 cP, respectively. We used Killough's hysteresis model for history-dependent capillary pressure and relative permeability functions (Killough 1976). The $CO_2$ viscosity was calculated based on the procedure described by Fenghour et al. (1998) and Vesovic et al. (1990). The mutual solubilities of $CO_2$ and $H_2O$ are calculated following the procedure given by Spycher and Pruess (2005), based on fugacity equilibration between water and a $CO_2$ phase. We calculated water fugacity using Henry's law and $CO_2$ fugacity using a modified Redlich-Kwong equation-of-state. We obtained gas density using a cubic equation-of-state tuned to accurately give the density of the compressed gas phase, following the procedure of Spycher and Pruess (2005). The diffusion flow in terms of liquid mole fraction is specified by the water phase diffusion coefficients for each component. The values of all three coefficients are $10^{-4}$ cm²/s. The gas phase diffusion coefficients for both water and $CO_2$ are $10^{-3}$ cm²/s.

To solve partial differential equations, we used the Adaptive IMplicit method (AIM), which is a compromise between the fully implicit and IMplicit Pressures Explicit Saturations (IMPES) method. The latter is similar to the fully implicit, except that all flow and well terms are computed using molar densities in a compositional run at the beginning of each time step. Note that IMPES scheme are quite accurate and they do not suffer from numerical dispersion (Moortgat et al, 2016). Time steps are chosen automatically.

### *3.3 Constitutive Relationships at fine scale for CG and FG rock types*

Here we present the constitutive relationships for the individual FG and CG rock types within a bar deposit. In the next section, the approach outlined in Section 2 is applied to average these and create the effective relationship for the bar deposit on whole. Multiple approaches have been developed to describe relative permeability-saturation-capillary pressure models (Burdine 1953; Corey 1954; Brooks and Corey 1964; Mualem 1976; van Genuchten 1980). Oostrom et al. (2016) showed that the Brooks and Corey (BC) model with the Corey-type relative permeability equations is the most suitable model for simulating supercritical $CO_2$ injection. This model that defines the individual characteristic curves for the individual FG and CG rock types in the drainage process. The capillary pressure $P_c$ is related to the brine (water) saturation $S_w$ by the equation (Brooks and Corey 1964):

$$P_c = P_e (S_w^*)^{-1/\lambda}, \tag{9}$$

where $S_w^* = \dfrac{S_w - S_{wi}}{1 - S_{wi}}$ is normalized water saturation, $S_{wi}$ is the irreducible water saturation, $P_e$ is the entry pressure, and $\lambda$ is the pore-size distribution index; $S_{CO2} = 1 - S_w$. The relations for the relative permeability for $CO_2$ and water are (Krevor et al. 2012) as follows:

$$k_{r,CO2} = k_{0,CO2} (1 - S_w^*)^2 (1 - (S_w^*)^{N_{CO2}}), \tag{10}$$

$$k_{r,w} = (S_w^*)^{N_w}, \qquad (11)$$

where $k_{0,CO2} = k_{CO2}(S_{wi})$, $N_{CO2}$ and $N_w$ are parameters that reflect the pore size distribution (Corey exponents).

For the imbibition, we used the Land approach (Land 1968), also described by Krevor et al. (2015). Only water-wet systems were considered. During imbibition, water replaces CO$_2$ and a part of the CO$_2$ becomes disconnected and trapped. The amount of residual (trapped) CO$_2$, $S_{CO2,r}$, is a function of the initial saturation before the beginning of imbibition, $S_{CO2,i}$:

$$S_{CO2,r}^* = \frac{S_{CO2,i}^*}{1 + C S_{CO2,i}^*}. \qquad (12)$$

where $C$ is the Land constant. The saturation of connected part of the CO$_2$ is

$$S_{CO2,c}^* = \frac{1}{2}[(S_{CO2}^* - S_{CO2,r}^*) + \sqrt{(S_{CO2}^* - S_{CO2,r}^*)^2 + \frac{4}{C}(S_{CO2}^* - S_{CO2,r}^*)}]. \qquad (13)$$

All imbibition curves are located between the drainage curve and bounding imbibition curve with maximal (normalized) residual saturation:

$$S_{CO2,r,max}^* = S_{CO2,r}^*(S_{CO2,i,max}^*) = \frac{1}{1+C}, \qquad (14)$$

where $S_{CO2,i,max}^* = \frac{S_{CO2,max}}{1-S_{w.i}} \equiv \frac{1-S_{w.i}}{1-S_{w.i}} = 1$. Thus, the maximal residual CO$_2$ is

$$S_{CO2,r,max} = \frac{1-S_{w.i}}{1+C} \qquad (15)$$

To define a bounding imbibition curve, we first find the saturation of the connected part of the CO$_2$ given maximal residual saturation. When combining (13) and (14), we get

$$S_{CO2,c,bound}^* = \frac{1}{2}[(S_{CO2}^* - \frac{1}{1+C}) + \sqrt{(S_{CO2}^* - \frac{1}{1+C})^2 + \frac{4}{C}(S_{CO2}^* - \frac{1}{1+C})}], \qquad (16)$$

Thus, imbibition curves for the capillary pressure and relative permeability for CO$_2$ are

$$k_{r,CO2} = k_{0,CO2} S^{*}_{CO2,c,bound}{}^{2} (1-(1-S^{*}_{CO2,c,bound})^{N_{CO2}}) \qquad (17)$$

$$P_c = P_e[(1-S^{*}_{CO2,c,bound})^{-1/\lambda} -1] \qquad (18)$$

Table 2 shows all the parameters necessary to create the capillary pressure and relative permeability curves for CG and FG rocks. Some of these values have been taken from experiments using samples of the Paaratte and Mt. Simon Sandstones (Krevor et al. 2012). The value of the Land constant was taken arbitrarily as $C = 1$ for both rock types. Then, $S_{CO2,r,max}$ was found using Eq. [15]. Figs. 6, 7, and 8 show the capillary pressure and relative permeability curves for CG and FG rocks for $CO_2$ and water.

**Table 2.** The values of parameters used for constitutive relations

|    | $S_{wi}$ | $P_e$ [Pa] | $\lambda$ | $k_{0,CO2}$ | $N_{CO2}$ | $N_w$ | $S_{CO2,r,max}$ | $k$[mD] | $\phi$ |
|----|------|------|------|------|-----|-----|-------|-----|-------|
| FG | 0.22 | 4600 | 0.55 | 0.94 | 4.0 | 9.0 | 0.39  | 11  | 0.244 |
| CG | 0.05 | 2100 | 0.9  | 0.95 | 2.0 | 8.0 | 0.475 | 112 | 0.283 |

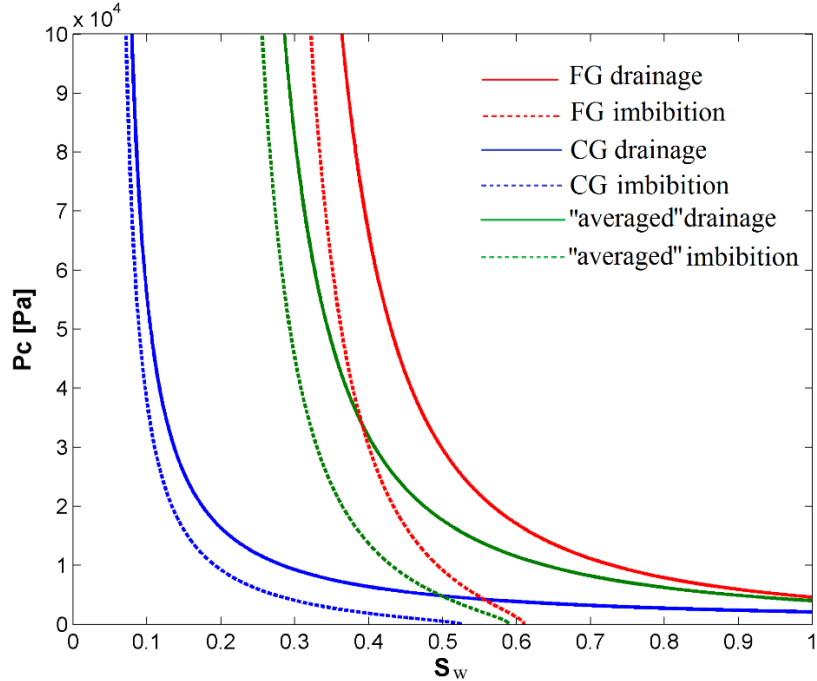

**Fig. 6.** Capillary pressure ($P_c$) as a function of brine saturation ($S_w$) in FG rocks (red) and CG rocks (blue) for drainage (solid curves) and imbibition (dashed curves). The green curves represent the effective capillary pressure curves "averaged" by the procedure described in the Methods section.

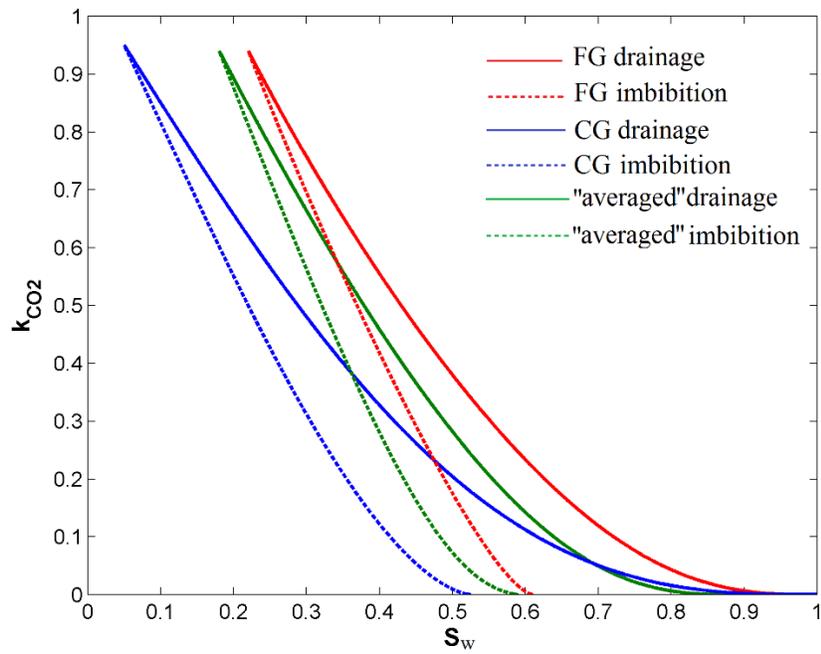

**Fig. 7.** Relative permeability for $CO_2$ ($k_{CO2}$) as a function of brine saturation ($S_w$) in FG rocks (red) and CG rocks (blue) for drainage (solid curves) and imbibition (dashed curves). The green curves represent the effective relative permeability curves "averaged" by the procedure described in the Methods section.

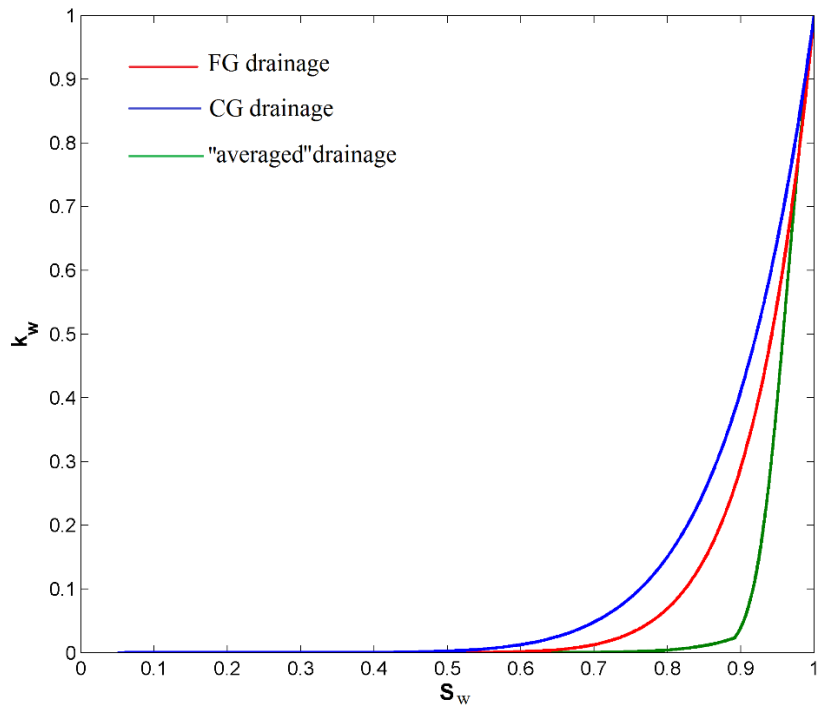

**Fig. 8** Relative permeability ($k_w$) for brine as a function of brine saturation (*S*) in FG rocks (red) and CG rocks (blue) for drainage. The green curve represents the effective relative permeability curve "averaged" by the procedure described in the Methods section.

### *3.4 Simulation with Effective Parameters and Relationships*

Using the approach detailed in Section 2 gives the following result for intrinsic permeability: $k_x = 36.7$ mD, $k_y = 22.0$ mD, and $k_z = 14.3$ mD, where the values of $k_{CG}$, $k_{CG}$, $h$, $H$, $l_x$ and $l_y$ are from the Table 1 and 2. The effective porosity is $\phi^{eff} = r_{FG}\phi_{FG} + r_{CG}\phi_{CG}$. The porosity is $\phi^{eff} = 0.2534$ for the parameter used (Table 2).

Using the approach detailed in Section 2 to define the effective capillary pressure curve gives the results in Fig. 6 (green color curves). Using the approach in Section 2 to define relative permeability gives the result in Figs. 7 and 8 (green color curves). The effective drainage curve is not merely a curve between FG and CG curves. Indeed, the values of effective relative permeability for CO$_2$ at a small CO$_2$ saturation (large water saturation) are smaller than the values of relative permeability for both FG and CG facies. The effective curve for the wetting phase is below both FG and CG curves over the entire saturation interval.

Three simulations were conducted in order to evaluate the approach. As a benchmark, Case (a) is a heterogeneous reservoir as shown in Figs 1 and 2 above, directly representing both the CG and FG cross sets using characteristic curves described in section 3 (Eqs. [9-18]; Table 2; Figs. 6–8). Both snap off trapping and capillary pinning are fully represented in Case (a).

Also, in order to compare the result using the effective relationships to a result without any capillary pinning, and thus to assess the importance of pinning, we also created Case (b) which is a homogeneous reservoir with anisotropic permeability (Eq. [6]) and characteristic curves given by FG facies.

Finally, Case (c) is a reservoir with anisotropic permeability (see Eq. [6]) and heterogeneity represented by the effective characteristic curves constructed using Eqs. [12–18] (Figs. 6-8). In the results presented in the next section, we show the spatial distribution of CO$_2$ saturation for these three reservoirs at the end of the injection stage (50 days) and after 1,000 days.

## 4. Results and Discussion

Comparing cases (a) and (b) as shown in panels a) and b) respectively in Fig. 9, the shape and the size of the plumes are similar during injection. This similarity indicates that "averaged" anisotropic permeability mostly imitates heterogeneous permeability map at a scale larger than the size of typical heterogeneity. The shape of the plume in case (c) is the same as in case (a), although the size is apparently a little bit smaller in the former. This difference is mostly because the plume in a heterogeneous reservoir has high permeability branches making the effective size larger. We hypothesize that this result could be improved if the code allowed defining anisotropic relative permeability. The effective relative permeability in the lateral direction would be more appropriately found by the arithmetic average, as with intrinsic permeability. Changing the code to allow this, and implementing it, would increase the width of the plume in case (c). However, this shortcoming is only evident during initial injection times. As shown next, the issue is not important when examining plume migration after injection.

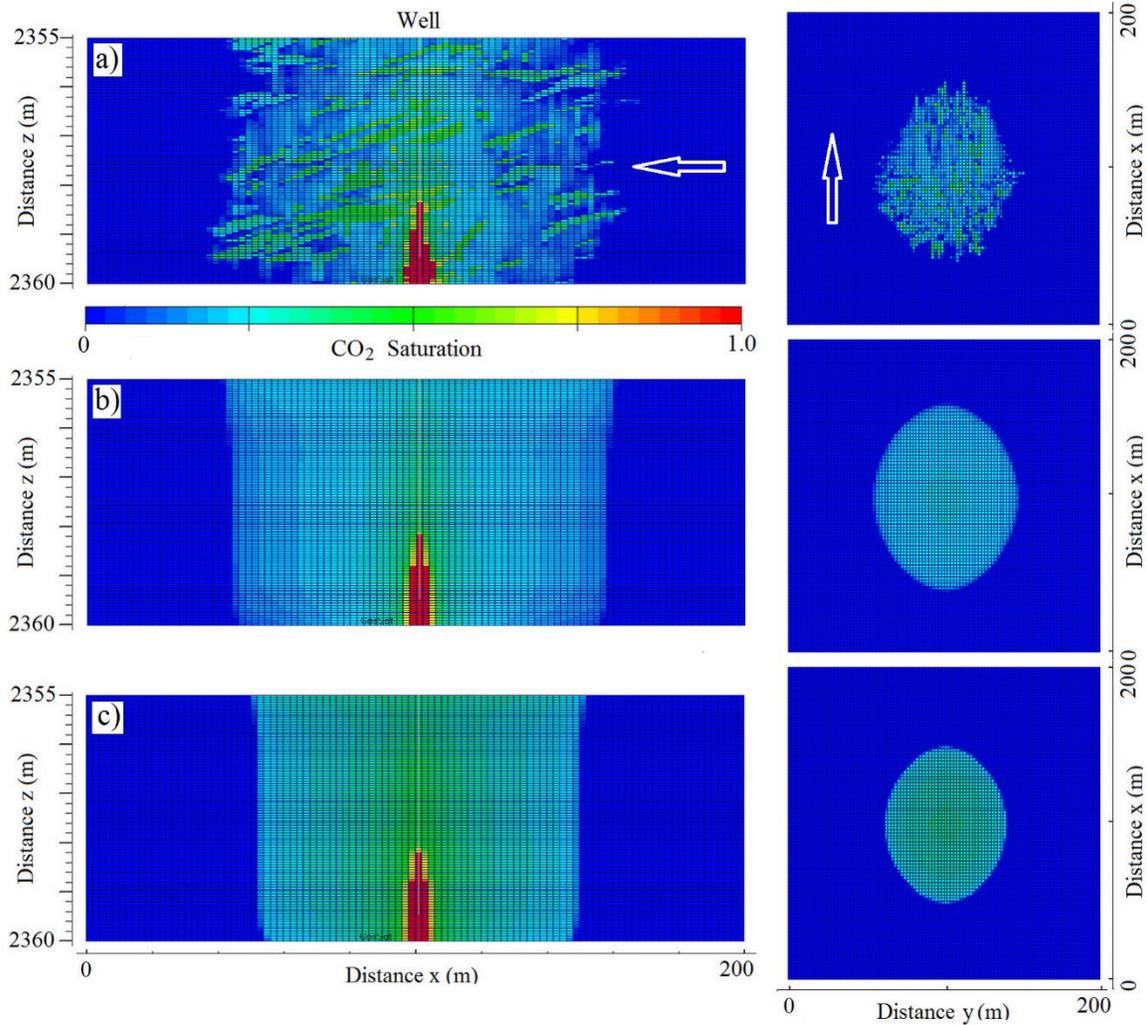

**Fig. 9.** Spatial distribution of $CO_2$ saturation in the middle cross-section of a reservoir (left panels) and aerial view (right panels) at the end of the injection stage (50 days) for a heterogeneous reservoir (a), homogeneous reservoir with anisotropic permeability (b), and reservoir with effective averaged parameters (c). Arrows show the fluvial direction. Vertical exaggeration of cross-section panels is 15×

During post-injection (Fig. 10), the difference in plume shape and size between cases (a) and (b) is more pronounced than during injection mainly because a larger portion of $CO_2$ migrates up in case (b) than in case (a). The latter happens because a homogeneous reservoir with "regular" (e.g., Brooks and Corey or van Genuchten type) characteristic curves includes only one type of capillary trapping, i.e. snap-off trapping. There are two types of capillary trapping in heterogeneous reservoir (case (a)), i.e., snap-off and capillary pinning. Capillary pinning is effectively included in a reservoir with effective characteristic curves representing smaller-scale heterogeneity. That is why there is virtually no difference in plume shape and size between cases (a) and (c) at Fig. 10. Comparison of cases (b) and (c) shows the obvious difference in trapping.

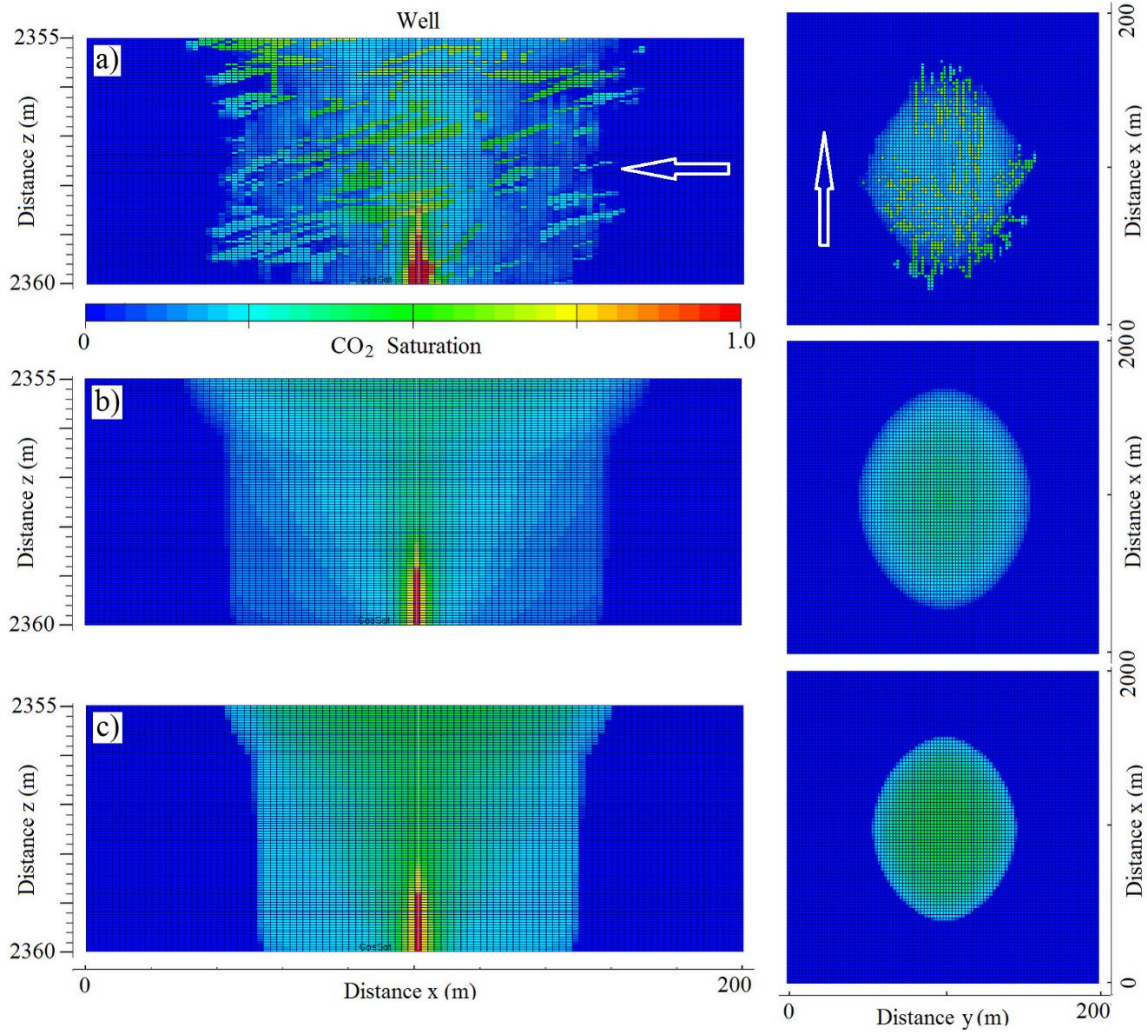

**Fig. 10** The same as at Fig. 9 after 1,000 days

## 5.0 Conclusion

Simulating $CO_2$ injection and redistribution in deep saline reservoirs is an important part of studying the $CO_2$ storage processes. Capillary trapping plays an essential role in the dynamics of the plume, given that trapped saturations could be up to 30% of the pore volume of the rock implied that even larger percentage of injected $CO_2$ becomes quasi-immobile almost immediately after injection (Krevor et al. 2015; Bachu 2015). Capillary trapping occurs because of heterogeneity at the pore scale (e.g., snap-off and pore doublets mechanisms (Chatzis and Dullien 1983; Hunt et al. 1988) and at scales from millimeters to the size of a reservoir (capillary pinning; Saadatpoor et al. 2010). The experimentally defined characteristic curves (relative permeability and capillary pressure) of most rock types describe relationships between parameters averaged for a volume much greater than the volume of an individual pore. The next step is to find the effective characteristic curves that describe the relationship between parameters averaged for a volume much greater than the size of typical heterogeneity. This study highlights an approach to describe this relationship for fluvial-type reservoirs with a typical size of heterogeneity (decimeter-meter).

Fluvial architecture causes $CO_2$ to propagate predominantly in the lateral direction, creating strong permeability anisotropy. Capillary pinning effect is strong in fluvial-type reservoirs because of a large difference between the capillary pressure of FG and CG sedimentary rock types (Gershenzon et al. 2014, 2015, 2016b, 2017). Thus, creating effective constitutive relations includes two steps: (1) finding an effective permeability tensor (Eq. [6]) and (2) finding effective capillary pressure and relative permeability curves by "averaging" the curves of FG and CG facies (Eqs. [4,5,7,8]). The latter procedure artificially includes the amount of capillary pinned $CO_2$ in a heterogeneous reservoir in the amount of snap-off $CO_2$ of a homogeneous reservoir. The benefits of averaging are improving iterative convergence and computational efficiency and thus the ability to (1) increase the overall size of the domain used in a reservoir simulation, and (2) to decrease simulation time by orders of magnitude.

Note that this approach is relevant in the context of determining integral amounts of $CO_2$ trapping. It is not relevant in the context of predicting first arrival of $CO_2$ away from the injection point. First arrival will be governed by the preferential flow pathways, including those created by connected, smaller scale, high permeability features which is not considered here. The effective saturation function approach presented in this article is not intended to solve first arrival problems, and would not be appropriate in that context.

The proposed approach has been given an initial evaluation in this article. These preliminary results showed that the approach represents snap-off trapping and capillary pinning in bar deposits reasonably well. Thus, the results suggest further evaluation is warranted, and the approach could be tested under a wider range of conditions and more comprehensive parameter space. One related issue that still needs work is the issue of dissolution. Our previous work has shown that small-scale heterogeneity significantly increases the surface area of the plume and thus promotes dissolution. Ideas for effective parameters controlling dissolution should be explored so that the increased dissolution trapping that occurs due to the heterogeneity is properly represented in simulations.

Summarizing from our recent studies (Gershenzon et al. 2014, 2015, 2016a, b, 2017), we may conclude that the risk of $CO_2$ storage in reservoirs with small-scale heterogeneity is less than in a homogeneous reservoir because of the additional capillary trapping processes arising from the heterogeneity, including capillary pinning and enhanced dissolution. The amount of trapped $CO_2$ by all processes depends on this heterogeneity as well as the injection rate. To maximize $CO_2$ trapping and determine optimal injection strategies (e.g. Bachu 2015; Dai et al. 2016) the small-scale heterogeneity in the permeable part of the reservoir should be represented in reservoir engineering models.

**Acknowledgments**

This work was supported as part of the Center for Geologic Storage of $CO_2$, an Energy Frontier Research Center funded by the U.S. Department of Energy, Office of Science, Basic Energy Sciences under Award # DE-SC0C12504. We acknowledge Schlumberger Limited for the donation of ECLIPSE Reservoir Simulation Software. This work was supported in part by the Ohio Supercomputer Center, which provided an allocation of computing time and technical support. We thank Albert Valocchi for useful comments and Daniel Klen for manuscript editing.

**Corrections for the article "Effective constitutive relations for simulating $CO_2$ capillary trapping in heterogeneous reservoirs with fluvial sedimentary architecture"**

In the article Gershenzon NI, Ritzi Jr RW, Dominic DF, and Mehnert E (2017), Effective Constitutive Relations for Simulating CO2 Capillary Trapping in Heterogeneous Reservoirs with Fluvial Sedimentary Architecture, Geomechanics and Geophysics for Geo-Energy and Geo-Resources, DOI 10.1007/s40948-017-0057-3 (hereafter G2017) several formulae should be corrected.

While presence of small-scale heterogeneity fundamentally affects dynamic of the plume, in particular $CO_2$ trapping, it is generally not computationally feasible to represent small-scale heterogeneity directly in a typical reservoir simulation. The goal is to include both types of capillary trappings, i.e., trapping on the pore-size level (snap-off) and on the level of small-scale heterogeneity (capillary pinning), into effective constitutive relations. Thus, the averaging volume should exceed not only the size of the pores but also the typical size of small-scale heterogeneity.

We suppose reservoir includes two rock types, course-grained (CG) and finer-grained (FG) (see Figs. 1 and 2 in G2017). The constitutive relations, i.e., capillary pressure and intrinsic and relative permeability, are different for these two rock types. Under quasi-equilibrium conditions, when viscous forces are much smaller than capillary forces, the following relation should be satisfied:

$$P_{c,CG}(S_{w,CG}) = P_{c,FG}(S_{w,FG}) - h\Delta\rho g, \qquad (1)$$

where $h\Delta\rho g$ is the buoyant force upward per area within this cross set, $\Delta\rho = \rho_w - \rho_{CO2}$; $\rho_w$ and $\rho_{CO2}$ are the density of brine and supercritical $CO_2$, respectively; $g$ is the gravitational constant, $h$ is a typical width of CG strata; $S_{w,CG}$ and $S_{w,FG}$ are the brine saturations and $P_{c,CG}$ and $P_{c,FG}$ are the capillary pressure of $CO_2$ for the CG and FG rock types.

The buoyant $CO_2$ moves from CG to FG strata then to CG strata and so on. In such a case the simplify conventional formula for the effective permeability (intrinsic permeability $k$ multiply by relative permeability $k_r$) is

$$k^{eff} k_{r,CO2}^{eff}(S_w^{eff}) = \frac{1}{r_{FG}/(k_{FG}k_{r,CO2,FG}(S_{w,FG})) + r_{CG}/(k_{CG}k_{r,CO2,CG}(S_{w,CG}))}, \qquad (2)$$

where $S_w^{eff} = S_{w,CG} r_{CG} + S_{w,FG} r_{FG}$, $r_{FG}$ and $r_{CG} = 1 - r_{FG}$ are the volume fractions of the FG and CG rocks, $k_{CG}$ and $k_{FG}$ are the intrinsic permeability of CG and FG rocks, and $k_{r,CO2,CG}$ and $k_{r,CO2,FG}$ are the relative permeability of $CO_2$ defined by the formulae (10) from the G2017 article. It is important to emphasize that saturations $S_{w,FG}$ and $S_{w,CG}$ obey relation (1). The effective intrinsic permeability is:

$$k^{eff} = \frac{1}{r_{CG}/k_{CG} + r_{FG}/k_{FG}}, \qquad (3)$$

From relations (2) and (3) we can find the effective relative permeability for $CO_2$ for drainage:

$$k_{r,CO2}^{eff}(S_{wi}^{eff} < S_w^{eff} < S_w^{eff,cr}) = \frac{r_{CG}/k_{CG} + r_{FG}/k_{FG}}{r_{FG}/(k_{FG}k_{r,CO2,FG}(S_{w,FG})) + r_{CG}/(k_{CG}k_{r,CO2,CG}(S_{w,CG}))},$$
$$k_{r,CO2}^{eff}(S_w^{cr,eff} < S_w^{eff} \leq 1) = 0, \qquad (4)$$

where $S_w^{eff,cr} = S_{w,CG}^{cr} r_{CG} + r_{FG}$ and $S_{w,CG}^{cr}$ satisfies the equation $P_{CG}^{cr}(S_{w,CG}^{cr}) = P_{e,FG} - h\Delta\rho g$ (see also Fig 1 for illustration). Note that the effective relative permeability of $CO_2$ is zero in the saturation range $S_w^{eff,cr} < S_w^{eff} \leq 1$, since $CO_2$ saturation is zero in the FG strata.

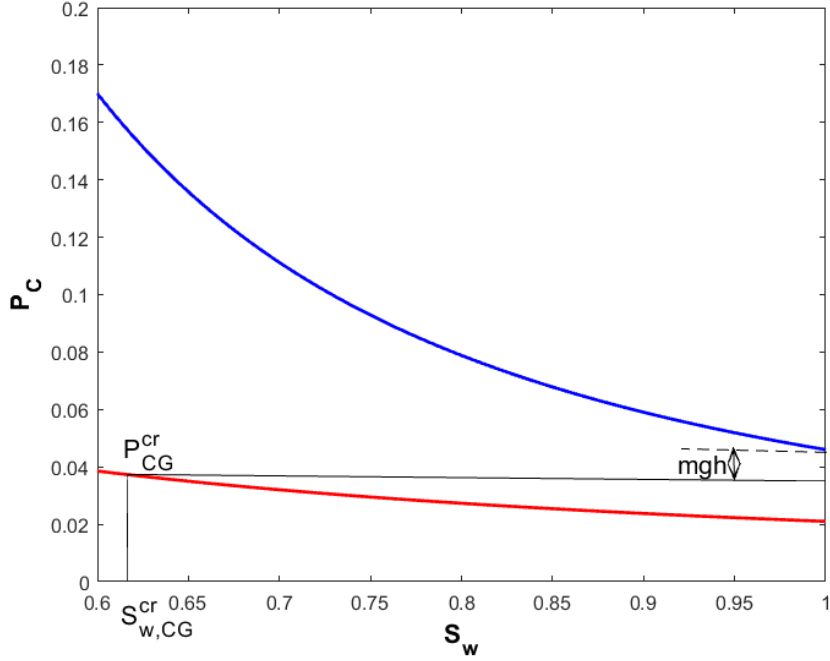

**Fig. 1** The schematic figure of capillary pressure (*P*c) for imbibition for FG (blue) and CG (red) facies as a function of water saturation (*S*w) illustrating the critical saturation $S_{w,CG}^{cr}$, $mg\Delta\rho$.

The effective relative permeability for brine is

$$k_{r,w}^{eff}(S_{wi}^{eff} < S_w^{eff} < S_w^{eff,cr}) = \frac{r_{CG}/k_{CG} + r_{FG}/k_{FG}}{r_{FG}/(k_{FG}k_{r,w,FG}(S_{w,FG})) + r_{CG}/(k_{CG}k_{r,w,CG}(S_{w,CG}))}$$

$$k_{r,w}^{eff}(S_w^{cr,eff} < S_w^{eff} \leq 1) = \frac{r_{CG}/k_{CG} + r_{FG}/k_{FG}}{r_{FG}/k_{FG} + r_{CG}/(k_{CG}k_{r,w,CG}(S_{w,CG}))}$$
(5)

where $k_{r,w,CG}$ and $k_{r,w,FG}$ are the relative permeability defined by the formulae (11) from the G2017 article. Fig. 2 shows the calculated effective relative permeability as well as the relative permeability for the CG and FG rocks.

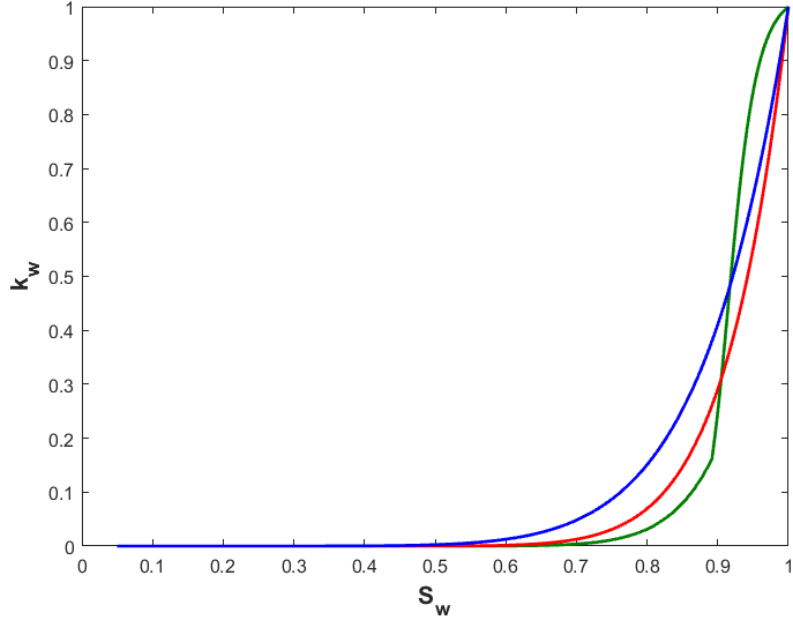

**Fig. 2** Relative permeability ($k_{r,w}$) for brine as a function of brine saturation ($S_w$) in FG rocks (red) and CG rocks (blue) for drainage calculated using formulae (11) from the G2017 article. The green curve represents the effective relative permeability curve "averaged" by the formula (5). The values of all necessary parameters are from the G2017 article. (For interpretation of the references to color in this figure legend, the reader is referred to the web version of this article.)

The effective capillary pressure for drainage is

$$P_c^{eff}(S_{wi}^{eff} < S_w^{eff} < S_w^{cr,eff}) = r_{FG} P_{c,FG}(S_{w,FG}) + r_{CG} P_{c,CG}(S_{w,CG})$$
$$P_c^{eff}(S_w^{cr,eff} < S_w^{eff} < 1) = r_{FG} P_{e,FG} + r_{CG} P_{c,CG}(S_{w,CG}) \tag{6}$$

To find the effective relative permeability and capillary pressure for imbibition we should first modify the formula (16) from G2017 for CG rocks by inclusion capillary pinning:

$$S_{CO2,c,bound,CG}^{*} = \frac{1}{2}[(S_{CO2,CG}^{*} - \frac{1}{1+C} - S_{pin}^{*}) + \sqrt{(S_{CO2,CG}^{*} - \frac{1}{1+C} - S_{pin}^{*})^2 + \frac{4}{C_{eff}}(S_{CO2,CG}^{*} - \frac{1}{1+C} - S_{pin}^{*})}],$$

(7)

where $S_{pin}^{*} = \frac{1 - S_{w,CG}^{cr}}{1 - S_{wi,CG}}$, $C_{eff} = \frac{1}{1/(1+C) + S_{pin}^{*}} - 1$, and $C$ is the Land constant. The formulae (17) and (18) from G2017 describe the imbibition curves for the capillary pressure and relative permeability for $CO_2$ for CG rocks, however variable $S_{CO2,c,bound}^{*}$ should be calculated using expression (7) from this article. The formulae for FG are the same as in G2017 article.

The effective capillary pressure and relative permeability for $CO_2$ for imbibition are

$$P_c^{eff}(S_{wi}^{eff} < S_w^{eff} < S_w^{eff,\max}) = r_{FG} P_{c,FG}(S_{w,FG}) + r_{CG} P_{c,CG}(S_{w,CG}) \qquad (8)$$

$$k_{r,CO2}^{eff}(S_{wi}^{eff} < S_w^{eff} < S_w^{eff,\max}) = \frac{r_{CG}/k_{CG} + r_{FG}/k_{FG}}{r_{FG}/(k_{FG}k_{r,CO2,FG}(S_{w,FG})) + r_{CG}/(k_{CG}k_{r,CO2,CG}(S_{w,CG}))}, \qquad (9)$$

where $S_w^{eff,\max} = r_{FG}\dfrac{1-S_{wi,FG}}{1+C} + r_{CG}(\dfrac{1-S_{wi,CG}}{1+C} + 1 - S_{w,CG}^{cr})$.

Figs 3 and 4 depict the respective effective curves calculated using formulae (4) - (9). The effective curves for imbibition 'absorb' capillary pinning effect.

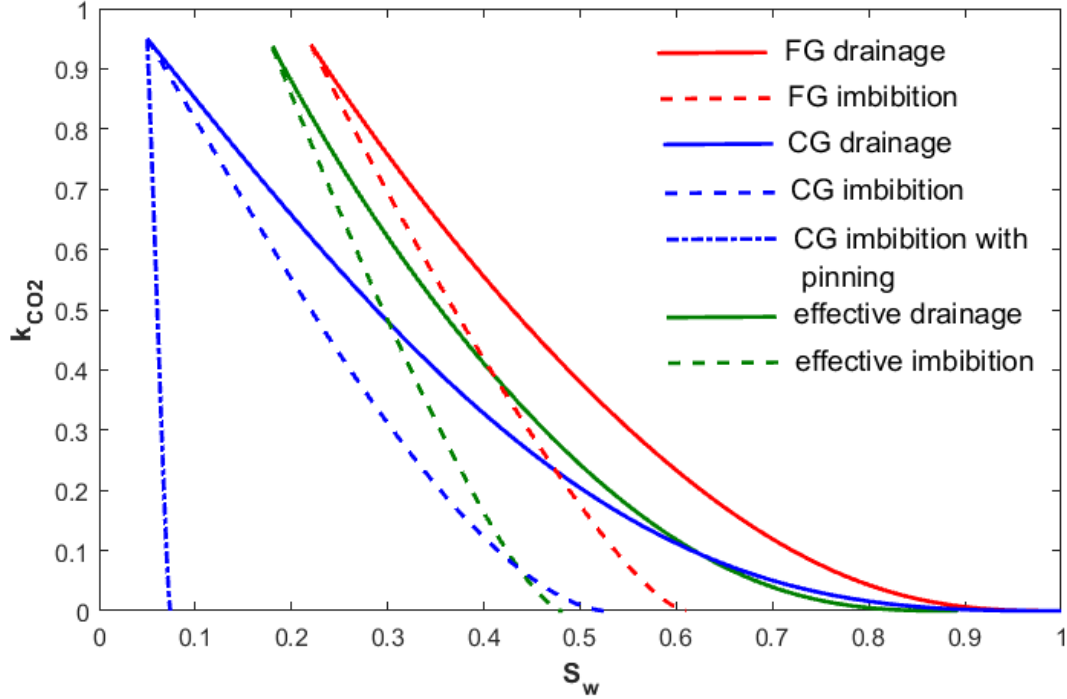

**Fig. 3** Relative permeability for $CO_2$ ($k_{CO2}$) as a function of brine saturation ($S_w$) in FG rocks (red) and CG rocks (blue) for drainage (solid curves) and imbibition (dashed curves). There are two curves for CG for imbibition, one is conventional (formulae (16) and (17) from the G2017 article) and the second one includes capillary pinning (formulae (17) from the G2017 article and (7) from this article). The green curves represent the effective relative permeability curves. (For interpretation of the references to color in this figure legend, the reader is referred to the web version of this article.)

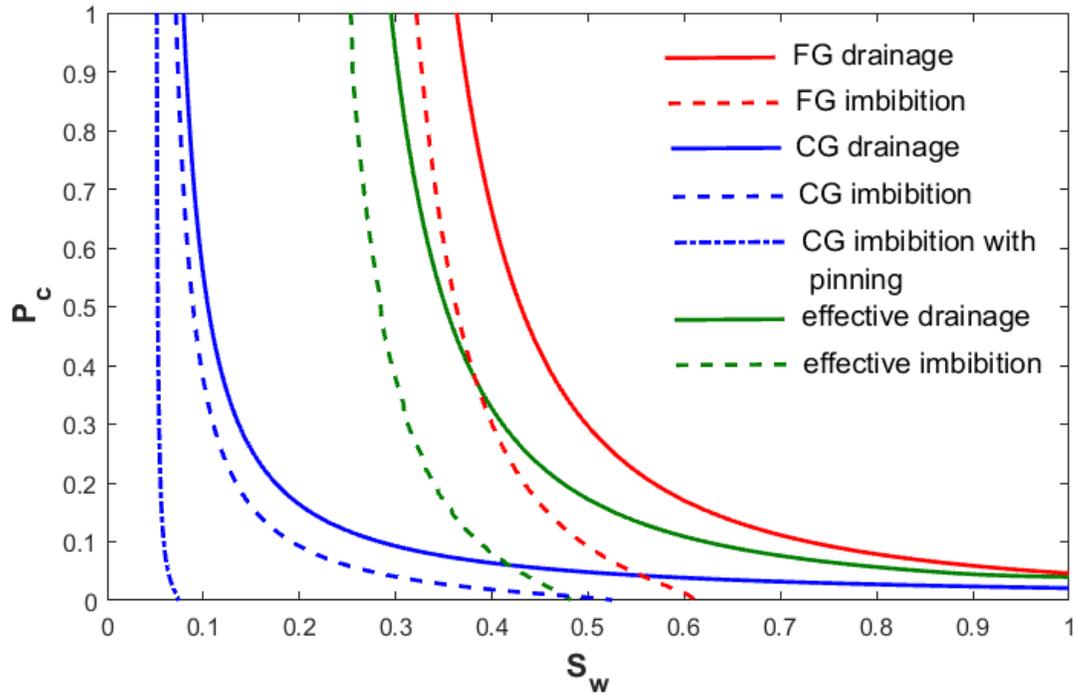

**Fig. 4** Capillary pressure ($P_c$) as a function of brine saturation ($S_w$) in FG rocks (red) and CG rocks (blue) for drainage (solid curves) and imbibition (dashed curves). There are two curves for CG for imbibition, one is conventional (formulae (16) and (18) from the G2017 article) and the second one includes capillary pinning (formulae (18) from the G2017 article and (7) from this article). The green curves represent the effective capillary pressure curves. (For interpretation of the references to color in this figure legend, the reader is referred to the web version of this article.)

Fig. 5 shows the spatial distribution of $CO_2$ saturation in the middle cross-section of a reservoir after 1000 days for a heterogeneous reservoir and homogeneous reservoir with effective parameters. As one can see, the shape of plume is similar in these two cases at large scale.

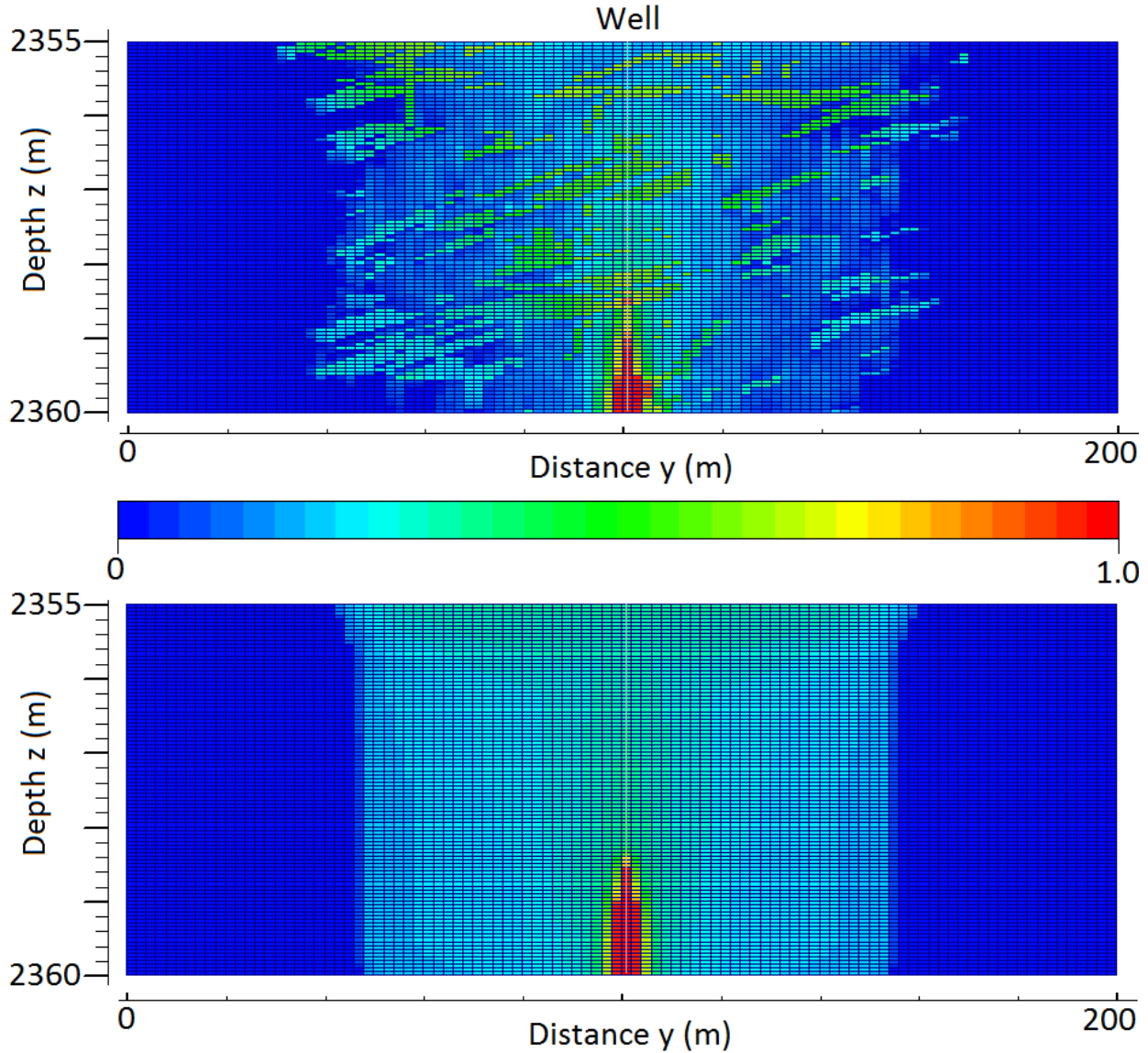

**Fig. 5** Spatial distribution of $CO_2$ saturation in the middle cross-section of a reservoir after 1000 days for a heterogeneous reservoir (upper panel) and homogeneous reservoir with effective parameters (bottom panel). Vertical exaggeration is 15×. Simulation parameters and condition are described in the G2017 article.

It is also useful to compare the amount of mobile and trapped $CO_2$ in these two cases (Fig. 6). The amount of capillary trapped $CO_2$ by snap-off mechanism in heterogeneous reservoir at injection stage (up to 50 days) is small compare to the amount of mobile $CO_2$ (compare the red and brown curves at Fig. 6, upper panel). However, as we can see, large part of this mobile $CO_2$ is not actually mobile due to capillary pinning effect (see blue and green curves at Fig. 6, upper panel). We already showed (Gershenzon et al 2017b) that at injection (i.e. drainage) stage the sizable amount of $CO_2$ is trapped by the capillary pinning. Here we can actually calculate this amount. After injection ceases the amount of $CO_2$ trapped by snap-off is largely increases (Fig. 6, bottom panel). The total amount of trapped $CO_2$ by both mechanisms is larger than the amount of mobile gas (compare blue and green curves at Fig. 6 bottom panel).

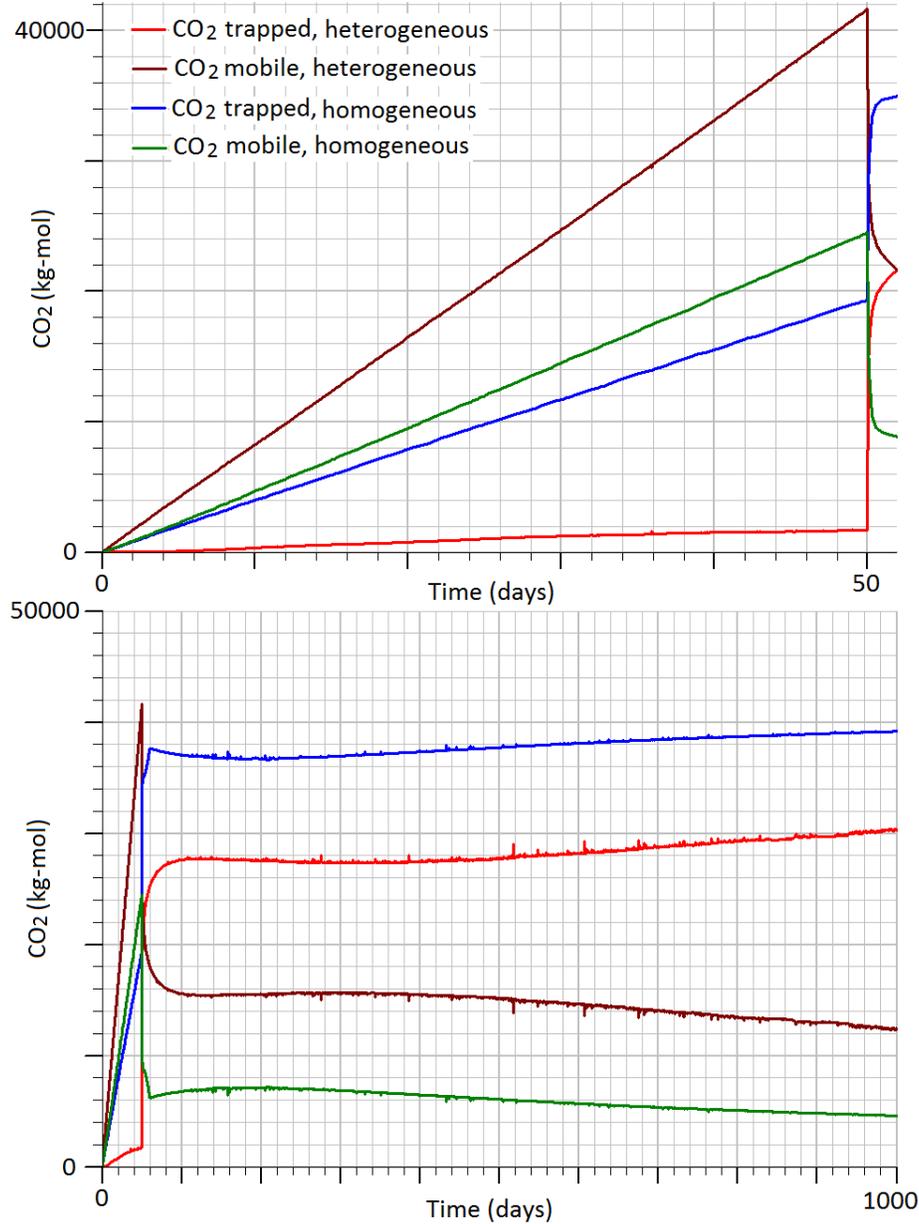

**Fig. 6** The total amount of trapped and mobile $CO_2$ as a function of time during CO2 injection period (upper panel) and injection and post-injection time (bottom panel) for the heterogeneous reservoir and homogeneous reservoir with effective constitutive relations. (For interpretation of the references to color in this figure legend, the reader is referred to the web version of this article.)

The maximal size of the cell in simulation of heterogeneous reservoir of fluvial type is limited by the typical size of heterogeneity, which is about 2 m in lateral direction and 5 cm in vertical direction. The size of the cell could be increased in homogeneous reservoir with effective constitutive relations, which allows simulation in realistic size reservoir. Figs. 7 and 8 show the results of simulation in reservoir with horizontal size of 2x2 km and vertical size of 50 m with injection rate of 100 kg/s during 2500 days (6.85 years). The total amount of injected $CO_2$ is 21.6 million tons. As in a previous simulation in a small size reservoir (compare Figs. 6 and 8), the total amount of trapped $CO_2$ is larger than the amount of mobile gas. Moreover, it is obvious that the further propagation of mobile $CO_2$ in the lateral direction results in reduction of mobile amount up to **complete $CO_2$ trapping by snap-off, pinning and dissolution**. Even if the improperly abandoned wells or stratigraphic discontinuities will locally compromise the integrity of caprock, the amount of $CO_2$ escaping reservoir will be negligible compare with the injected amount.

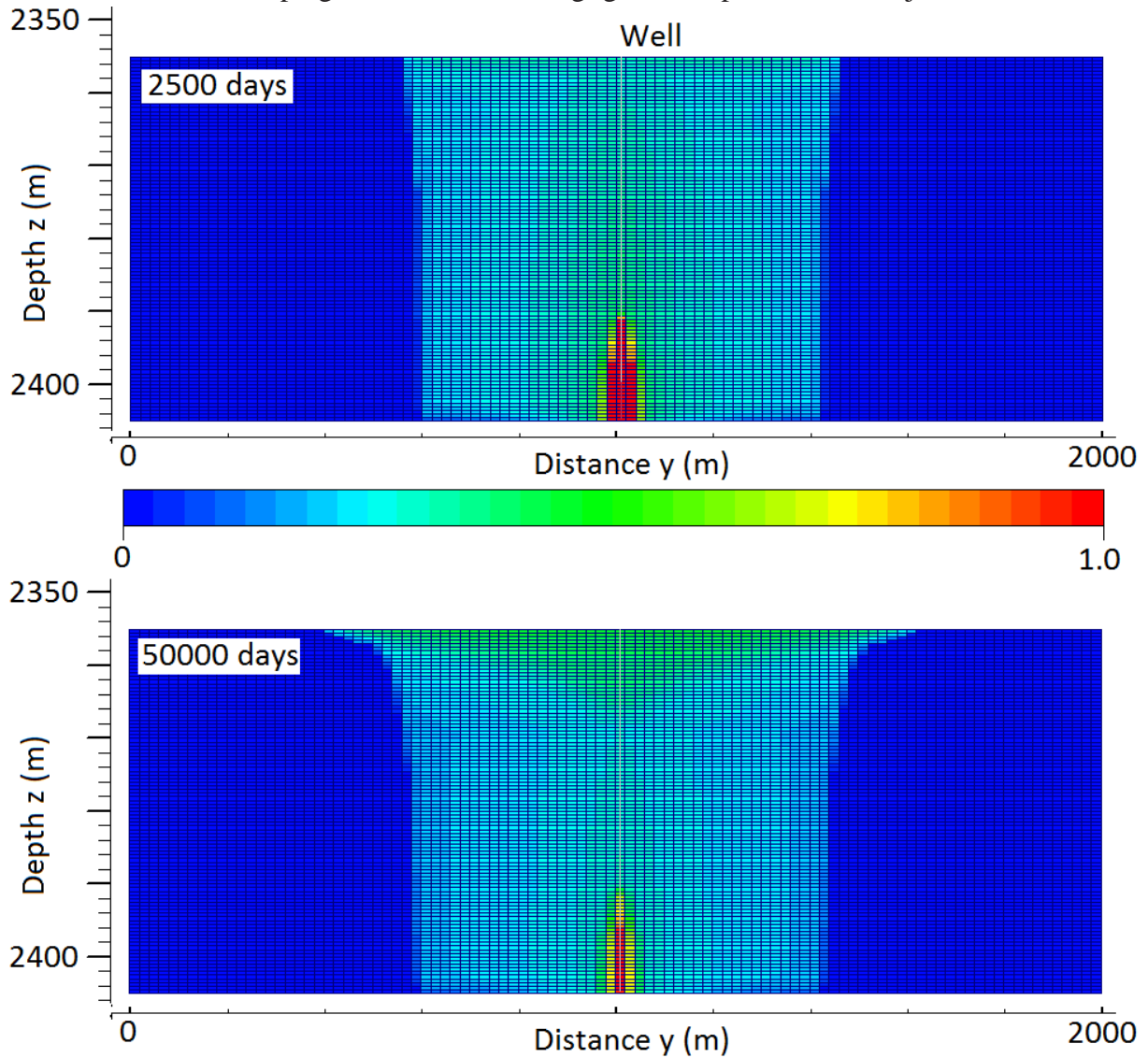

**Fig. 7** Spatial distribution of $CO_2$ saturation in the middle cross-section of a reservoir immediately after injection (2500 days = 6.85 years) and after 50000 days (137 years). Vertical exaggeration is 15×.

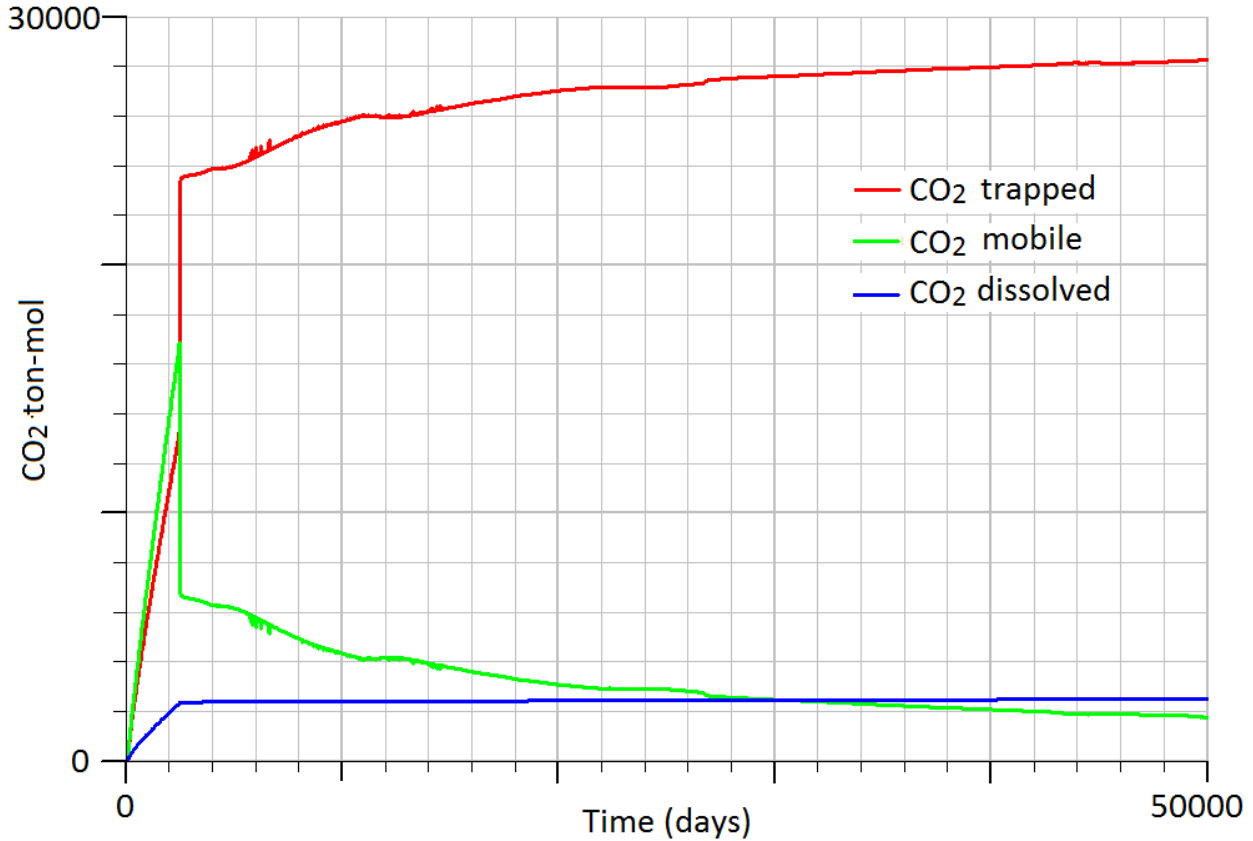

**Fig. 8** The total amount of trapped, mobile CO$_2$, and dissolved as a function of time homogeneous reservoir with effective constitutive relations. (For interpretation of the references to color in this figure legend, the reader is referred to the web version of this article.)

Here we derive the formulae for the effective constitutive relations describing supercritical CO$_2$ bouncy in heterogeneous reservoir with fluvial sedimentary architecture. This approach allows simulation of CO$_2$ sequestration in realistic size reservoir and reduces simulation time by at one to two orders of magnitude.